\documentclass[
aps,%
12pt,%
final,%
notitlepage,%
oneside,%
onecolumn,%
nobibnotes,%
nofootinbib,%
superscriptaddress,%
showpacs,%
centertags]%
{revtex4}

\def\'#1{{\accent19\ifx #1i \i\else #1\fi}}

\def\jour#1#2#3#4{{\sl #1{}} {\bf #2}, #3\ (#4)}

\newcommand{\be}{\begin{equation}}
\newcommand{\ee}{\end{equation}}
\newcommand{\bea}{\begin{eqnarray}}
\newcommand{\eea}{\end{eqnarray}}

\begin{document}
\selectlanguage{english}
\title{Contractions and deformations of quasi-classical Lie algebras
preserving a non-degenerate quadratic Casimir operator }
\author{\firstname{ Rutwig}~\surname{Campoamor-Stursberg}}
\email{rutwig@mat.ucm.es} \affiliation{Dpto. Geometr\'{\i}a y
Topolog\'{\i}a, Fac. CC. Mat., Universidad Complutense, Plaza de
Ciencias 3, 28040 Madrid, Spain }

\begin{abstract}
By means of contractions of Lie algebras, we obtain new classes of
indecomposable quasi-classical Lie algebras that satisfy the
Yang-Baxter equations in its reformulation in terms of triple
products. These algebras are shown to arise naturally from
non-compact real simple algebras with non-simple complexification,
where we impose that a non-degenerate quadratic Casimir operator
is preserved by the limiting process.  We further consider the
converse problem, and obtain sufficient conditions on integrable
cocycles of quasi-classical Lie algebras in order to preserve
non-degenerate quadratic Casimir operators by the associated
linear deformations.
\end{abstract}
 \maketitle

\section{Introduction: }

Reductive Lie algebras have been shown to be the most convenient
class of algebras for physical applications. They arise naturally
as the Lie algebras of compact groups, and contain the class of
semisimple algebras. Moreover, they have an important property,
namely an invariant metric\footnote{This metric arises immediately
from the Killing tensor for the semisimple case.}, of crucial
importance in problems like defining Wess-Zumino-Witten models.
Classically made on semisimple and reductive algebras, models
based on non-reductive algebras have been shown to be of physical
interest \cite{Wi}. Other important applications of Lie algebras
endowed with a symmetric non-degenerate invariant form, which we
call here quasi-classical\footnote{Other authors call algebras
like these symmetric self-dual.}, are for example conformal field
theory, where they correspond to the Lie algebras admitting a
Sugawara construction, or the Yang-Baxter equations, where
quasi-classical algebras provide classes of solutions
\cite{Fi,Dr,Ok}.

\noindent

A Lie algebra $L$ is called quasi-classical if it possesses a
bilinear symmetric form $\langle .,.\rangle$ that satisfies the
constraints \bea \langle \left[X,Y\right],Z \rangle = \langle
X,\left[Y,Z\right]\rangle, \\
{\rm If} \langle X,Y\rangle=0, \forall X\in L \Rightarrow Y=0.
\eea

The first condition shows that the bilinear form satisfies an
associativity condition (also called invariance), while the second
expresses non-degenerateness. Given a basis
$\left\{X_{1},..,X_{n}\right\}$ of $L$ and the corresponding
structure tensor $\left\{C_{ij}^{k}\right\}$, we obtain the
expression of $\langle .,.\rangle$ as: \be \langle X_{i},X_{j}
\rangle= g_{ij}. \ee

Since the form is non-degenerate, we find an inverse to the
coefficient matrix of $\langle .,.\rangle$:
$g^{ij}=(g_{ij})^{-1}$. Obviously semisimple Lie algebras satisfy
these requirements for the Killing form. Also reductive and
abelian Lie algebras are trivially quasi-classical although in
this case the Killing metric is no more non-degenerate. In
\cite{Ok} it was shown that a necessary and sufficient condition
for the existence of such a form is that $L$ admits a
non-degenerate quadratic Casimir operator
$C=g^{\alpha\beta}x_{\alpha}x_{\beta}$. Using the realization by
differential operators
$\widehat{X}_{i}=C_{ij}^{k}x_{k}\frac{\partial}{\partial_{x_{j}}}$,
this means that $C$ is a solution of the following system of
partial differential equations: \be
C_{ij}^{k}x_{k}\frac{\partial}{\partial_{x_{j}}}C=0. \ee

Using this characterization, we obtain a useful criterion to test
whether a Lie algebra is quasi-classical or not, and in certain
situations more practical than various pure algebraic structural
results (see e.g. \cite{Fa} and references therein). In
particular, for any given dimension, the classification of
quasi-classical Lie algebras follows from the classification of
isomorphism classes once the invariants of the coadjoint
representation have been computed. Therefore the problem of
finding metrics reduces to an analytical problem, which is solved
in low dimension \cite{Pa,C48}.

\medskip

This paper is structured as follows: In section 2 we reformulate
the Yang-Baxter equations in terms of triple products, which
enables us to obtain some sufficiency criteria basing only on the
structure tensor of a quasi-classical algebra.  This triple
product formulation is used in combination with contractions of
Lie algebras to construct large classes of indecomposable
quasi-classical algebras that preserve the quadratic
non-degenerate Casimir operator of a semisimple classical Lie
algebra. In section 3 we focus on a kind of inverse problem,
namely, deformations of quasi-classical Lie algebras that preserve
the quadratic Casimir operator, and therefore, the associated
metric. This leads to a characterization of such deformations in
terms of integrable cocycles in the adjoint cohomology.

\section{Yang-Baxter equations and quasi-classical algebras}

\subsection{Yang-Baxter equations and triple products}

Yang-Baxter equations (YBE) have been known a long time to embody
the symmetries of two dimensional integrable models \cite{Ji}, and
also appear in many problems concerning statistical physics and
quantum groups. In addition to the classical semisimple case,
non-reductive quasi-classical Lie algebras were recognized to
provide some solutions of the YBE when these are rewritten in
terms of triple products \cite{Ok3}. With this reformulation, some
useful sufficient conditions can be found in dependence of the
structure tensor of the quasi-classical Lie algebra.

Given a finite dimensional vector space $V$ with inner product
$\left\langle .,.\right\rangle $, then, for any basis $\left\{
v_{1},..,v_{n}\right\}$, we set the coefficients in the usual way
\begin{equation*}
\langle v_{i},v_{j}\rangle :=g_{ij}=g_{ji}
\end{equation*}
and define the raising of indices
\[
v^{j}=\sum_{i=1}^{n}g^{ij}v_{i}.
\]
Given a spectral parameter $\theta,$ we consider the map $R\left(
\theta\right)  :V\otimes V\rightarrow V\otimes V$ defined by
\be
R\left(  \theta\right)  \left(  v_{i}\otimes v_{j}\right)  =\sum_{k,l=1}%
^{n}R_{ij}^{kl}\left(  \theta\right)  v_{k}\otimes v_{l}. \ee We
obtain the Yang-Baxter equations in its usual form \cite{Ji} from
the relations \be R_{12}\left(  \theta\right)  R_{13}\left(
\theta^{\prime}\right) R_{23}\left(  \theta^{\prime\prime}\right)
=R_{23}\left(  \theta ^{\prime\prime}\right)  R_{13}\left(
\theta^{\prime}\right) R_{12}\left( \theta\right)  ,
\ee where
$\theta^{\prime\prime}=\theta^{\prime}-\theta$. The equations can
be rewritten using triple products, which provides sometimes a
more convenient
presentation for solutions governed by certain types of purely solvable
quasi-classical Lie algebras. Introducing the triple products \cite{Ok3}%
\be
\left[  v^{j},v_{k},v_{l}\right]
_{\theta}=\sum_{i=1}^{n}R_{kl}^{ij}\left(
\theta\right)  v_{i},\;\left[  v^{i},v_{k},v_{l}\right]  _{\theta}^{\ast}%
=\sum_{j=1}^{n}R_{kl}^{ij}\left(  \theta\right)  v_{i},
\ee
the YBE reduces to the relation%
\be \sum_{j=1}^{n}\left[  u,\left[  v,v_{j},w\right]
_{\theta^{\prime}},\left[ e^{j},x,y\right]  _{\theta}\right]
_{\theta^{\prime\prime}}^{\ast}=\sum _{j=1}^{n}\left[  v,\left[
u,v_{j},x\right]  _{\theta^{\prime}},\left[ e^{j},w,y\right]
_{\theta^{\prime\prime}}^{\ast}\right]  _{\theta}, \label{YB11}
\ee
where $u,v,w,x,y\in V$. A particularly interesting case is given
when the scattering matrix elements $R_{kl}^{ij}\left(
\theta\right)  $ satisfy the
following constraint:%
\be R_{kl}^{ij}\left(  \theta\right)  -R_{lk}^{ji}\left(
\theta\right)  =0.
\ee

In this case, the equation (\ref{YB11}) becomes%
\be
 \sum_{j=1}^{n}\left[  u,\left[  v,v_{j},w\right]
_{\theta^{\prime}},\left[
e^{j},x,y\right]  _{\theta}\right]  _{\theta^{\prime\prime}}=\sum_{j=1}%
^{n}\left[  v,\left[  u,v_{j},x\right]  _{\theta^{\prime}},\left[
e^{j},w,y\right]  _{\theta^{\prime\prime}}\right]  _{\theta},
\ee
subjected to the condition
\[
\left\langle u,\left[  v,w,x\right]  _{\theta}\right\rangle
=\left\langle v,\left[  u,x,w\right]  _{\theta}\right\rangle .
\]
Even in this case, the solving of the equations is far from being
trivial. However, it was found in \cite{Ok3} that if $L$ satisfies the condition%
\begin{equation}
\left[  L,\left[  \left[  L,L\right]  ,\left[  L,L\right]  \right]
\right]
=0,\label{MA}%
\end{equation}
then we have commutation relation \be \left[  R_{jk}\left(
\theta\right) ,R_{lm}\left( \theta^{\prime}\right) \right]
=0,\;j,k,l,m=1,2,3. \label{MA1} \ee This is particular implies
that the YBE (\ref{YB11}) is satisfied. Classes of solvable Lie
algebras in arbitrary dimension that satisfy these conditions have
been constructed in \cite{Ok3}, as well as examples where the
commutation relation (\ref{MA1}) is not necessarily satisfied.

\subsection{Contractions of quasi-classical Lie algebras}

In this paragraph we obtain additional classes of indecomposable
nilpotent Lie algebras satisfying condition  (\ref{MA}). In
comparison with previous constructions, the class of algebras
obtained here follows naturally from contractions of simple real
Lie algebras that preserve the quadratic Casimir operator. We can
therefore construct quasi-classical algebras with prescribed inner
product.

We recall that a contraction $L\rightsquigarrow L^{\prime}$ of a
Lie algebra is given by the commutators
\begin{equation}
\left[X,Y\right]^{\prime}:=\lim_{t\rightarrow \infty}
\Phi_{t}^{-1}\left[\Phi_{t}(X),\Phi_{t}(Y)\right], \label{Kow}
\end{equation}
where $\Phi_{t}$ is a parameterized family of non-singular linear
maps in $L$ for all $t<\infty$.\footnote{It is assumed that the
limit (\ref{Kow}) exists for any pair $X,Y$ of generators in $L$.}
Among the various types of contractions existing, we consider here
the so-called generalized In\"on\"u-Wigner contractions given by
automorphims of the type\footnote{For properties of this type of
contractions see e.g. \cite{We}.}
\begin{equation}
\Phi_{t}(X_{i})=t^{-n_{i}}X_{i},\quad n_{i}\in\mathbb{Z}.
\end{equation}
Contractions can also be extended to invariants. Let
$F(X_{1},...,X_{n})=\alpha^{i_{1}...i_{p}}X_{i_{1}}...X_{i_{p}}$
be a Casimir operator of degree $p$. Expressing it over the
transformed basis we get
\begin{equation}
F(\Phi_{t}(X_{1}),..,\Phi_{t}(X_{n}))=t^{n_{i_{1}}+...+n_{i_{p}}}\alpha^{i_{1}...i_{p}}X_{i_{1}}...X_{i_{p}}.
\end{equation}
Now let
\begin{equation}
M=\max \left\{n_{i_{1}}+...+n_{i_{p}}\quad |\quad
\alpha^{i_{1}..i_{p}}\neq 0\right\},
\end{equation}
and consider the limit
\begin{equation}
F^{\prime}(X_{1},..,X_{n})=\lim_{t\rightarrow \infty}
t^{-M}F(\Phi_{t}(X_{1}),...,\Phi_{t}(X_{n}))=\sum_{n_{i_{1}}+...+n_{i_{p}}=M}
\alpha^{i_{1}...i_{p}}X_{i_{1}}...X_{i_{p}}.
\end{equation}
It is not difficult to see that this expression is  a Casimir
operator of degree $p$ of the contraction. Imposing that the
invariant remains unchanged by the contraction implies certain
restriction that must not necessarily occur \cite{C43}. For our
purpose, preservation of a non-degenerate quadratic Casimir
operator implies automatically that the contraction is
quasi-classical, and the induced inner product the same. To this
extent, let $\frak{s}$ be a complex semisimple Lie algebra of
classical type $A_{l},B_{l},C_{l},D_{l}$ and let $\frak{sl}\left(
l+1,\mathbb{C}\right)
,\frak{so}\left(  2l+1,\mathbb{C}\right)  ,\frak{sp}\left(  2l,\mathbb{C}%
\right)  \,$\ and $\frak{so}\left(  2l,\mathbb{C}\right)  $ be the
non-compact simple real Lie algebra with complexification
$\frak{s\oplus s}$. Let $\left\{  X_{1},..,X_{n}\right\}  $ and
$\left\{  Y_{1},..,Y_{n}\right\}  $ be a basis of each copy of
$\frak{s}$ such that
\[
\left[  X_{i},X_{j}\right]  =C_{ij}^{k}X_{k},\;\left[
Y_{i},Y_{j}\right] =C_{ij}^{k}Y_{k},\;\left[  X_{i},Y_{j}\right]
=0,
\]
i.e., the structure tensor is the same in both copies. Considering
the change of basis given by
\be
\overline{X}_{i}=X_{i}+Y_{i},\;\overline{Y}_{i}=\sqrt{-1}\left(  Y_{i}%
-X_{i}\right)  ,\;i=1,..,n,\; \ee the structure tensor over the
basis $\left\{ \overline{X}_{1},..,\overline
{X}_{n},\overline{Y}_{1},..,\overline{Y}_{n}\right\}  $ is
expressed by
\be
\left[  \overline{X}_{i},\overline{X}_{j}\right]  =C_{ij}^{k}\overline{X}%
_{k},\;\left[  \overline{X}_{i},\overline{Y}_{j}\right]  =C_{ij}^{k}%
\overline{Y}_{k},\;\left[ \overline{Y}_{i},\overline{Y}_{j}\right]
=-C_{ij}^{k}\overline{X}_{k}.
\ee
Since $\frak{s}\oplus\frak{s}$
is quasi-classical for being semisimple, it
admits quadratic Casimir operators%
\[
C_{1}=g^{ab}X_{a}X_{b},\;C_{2}=g^{ab}Y_{a}Y_{b}.
\]
Suitable linear combinations of them provide a nondegenerate
quadratic operator on the direct sum. Rewriting these operators in
the new basis, we obtain the operators \be
\overline{C}_{1}=g^{ij}\left(
\overline{X}_{i}\overline{X}_{j}-\overline
{Y}_{i}\overline{Y}_{j}\right)  ,\;\overline{C}_{2}=g^{ij}\left(
\overline
{X}_{i}\overline{Y}_{j}+\overline{Y}_{i}\overline{X}_{j}\right)  ,
\ee
 which are easily seen to be non-degenerate. It is natural to
ask whether there exist non-trivial contractions of
$\frak{s}\oplus\frak{s}$ such that the contraction preserves at
least one of the preceding Casimir operators. In this way, the
contraction is also quasi-classical with the same bilinear form as
the contracted algebra. The suitable operator to be tested is
$\overline{C}_{2}$. There is the well known obvious contraction
$\frak{s}\oplus \frak{s\rightsquigarrow
s}\overrightarrow{\oplus}_{ad\left(  \frak{s}\right)
}nL_{1}$ determined by the parameterized changes of basis%
\[
F_{t}\left(  \overline{X}_{i}\right)
=\overline{X}_{i},\;F_{t}\left( \overline{Y}_{i}\right)
=\frac{1}{t}\overline{Y}_{i},\;i=1,..,n.
\]
The contracted invariants are as follows:%
\bea
\lim_{t\rightarrow\infty}\frac{1}{t^{2}}\overline{C}_{1}   =-g^{ij}%
\overline{Y}_{i}\overline{Y}_{j},\nonumber \\
\lim_{t\rightarrow\infty}\frac{1}{t}\overline{C}_{2} =g^{ij}\left(
\overline{X}_{i}\overline{Y}_{j}+\overline{Y}_{i}\overline{X}_{j}\right)
. \eea
Therefore the contraction preserves the invariant
$\overline{C}_{2}$ and is
quasi-classical. There is another possibility of contracting $\frak{s}%
\oplus\frak{s}$ that also leads to a quasi-classical contraction.
Consider the parameterized change of basis
\[
G_{t}\left(  \overline{X}_{i}\right)  =\frac{1}{t}\overline{X}_{i}%
,\;G_{t}\left(  \overline{Y}_{i}\right)  =\frac{1}{\sqrt{t}}\overline{Y}%
_{i},\;i=1,..,n.
\]
Then the contraction $L^{\prime}$ has the following brackets%
\be
 \left[  \overline{X}_{i},\overline{X}_{j}\right]  =0,\;\left[
\overline {X}_{i},\overline{Y}_{j}\right]  =0,\;\left[
\overline{Y}_{i},\overline {Y}_{j}\right]
=-C_{ij}^{k}\overline{X}_{k}.
\ee
 Therefore $L^{\prime}$ is an indecomposable
nilpotent Lie algebra with $n$-dimensional centre. Contracting the
invariants of $\frak{s}\oplus\frak{s}$ we obtain
\bea
\lim_{t\rightarrow\infty}\frac{1}{t^{2}}\overline{C}_{1}   =g^{ij}%
\overline{X}_{i}\overline{Y}_{j},\nonumber \\
\lim_{t\rightarrow\infty}\frac{1}{t}\overline{C}_{2} =g^{ij}\left(
\overline{X}_{i}\overline{Y}_{j}+\overline{Y}_{i}\overline{X}_{j}\right)
. \eea
Thus the contraction is again quasi-classical, and it further satisfies the conditions%
\[
C^{1}L^{\prime}=\left[  L^{\prime},L^{\prime}\right]
=\left\langle \overline{X}_{1},..,\overline{X}_{n}\right\rangle
,\;C^{2}L^{\prime}=\left[ L^{\prime},C^{1}L^{\prime}\right]  =0,
\]
showing that $L^{\prime}$ is a metabelian Lie algebra. It actually
satisfies the condition (\ref{MA}), and therefore the Yang-Baxter
equations (\ref{YB11}). The same formal construction holds for
direct sums of copies of exceptional complex Lie algebras.

Using specific realizations, like boson and fermion operators,
other classes of quasi-classical algebras that preserve the
quadratic Casimir operator of a simple Lie algebra can be
constructed \cite{C43}. This occurs for example for the symplectic
Lie algebras $\frak{sp}(N,\mathbb{R})$ given by the creation and
annihilation operators: consider the linear operators
$a_{i},a_{j}^{\dagger}\;\left(  i,j=1..N\right)  $ satisfying the
commutation relations \be \left[  a_{i},a_{j}^{\dagger}\right]
=\delta_{ij}\mathbb{I},\quad \left[  a_{i},a_{j}\right]    =\left[
a_{i}^{\dagger},a_{j}^{\dagger }\right]  =0 \label{Kl2} \ee
Considering the operators $\left\{  a_{i}^{\dagger}a_{j},a_{i}^{\dagger}%
a_{j}^{\dagger},a_{i}a_{j}\right\}  $,  we generate
$\frak{sp}\left( N,\mathbb{R}\right)  $, the brackets of which
follow easily from (\ref{Kl2}). If we introduce the labelling \be
X_{i,j}    =a_{i}^{\dagger}a_{j},\quad X_{-i,j}
=a_{i}^{\dagger}a_{j}^{\dagger},\quad X_{i,-j} =a_{i}a_{j},\;1\leq
i,j\leq N \label{Er2}, \ee
the brackets of $\frak{sp}\left(
N,\mathbb{R}\right) $ are comprised as: \be
\left[  X_{i,j},X_{k,l}\right]  =\delta_{jk}X_{il}-\delta_{il}X_{kj}%
+\varepsilon_{i}\varepsilon_{j}\delta_{j,-l}X_{k,-i}-\varepsilon
_{i}\varepsilon_{j}\delta_{i,-k}X_{-j,l}, \label{Kl3} \ee where
$-N\leq i,j,k,l\leq N$, $\varepsilon_{i}={\rm sgn}\left( i\right)
$ and $X_{i,j}+\varepsilon_{i}\varepsilon_{j}X_{-j,-i}=0$.
Defining now the family of automorphisms \bea
\Phi_{t}(X_{i,i})=\frac{1}{\sqrt{t}}X_{i,i},\; (1\leq i\leq
N),\quad \Phi_{t}(X_{i,j})=\frac{1}{t}X_{i,j},\; (i < j),\quad
\Phi_{t}(X_{i,j})=X_{i,j},\; (i > j)\nonumber\\
\Phi_{t}(X_{-i,j})=\frac{1}{t}X_{-i,j},\quad
\Phi_{t}(X_{i,-j})=X_{i,-j},\; (1\leq i,j\leq N), \eea it is
straightforward to verify that
\begin{equation}
\lim_{t\rightarrow \infty}
\Phi_{t}^{-1}\left[\Phi_{t}(X),\Phi_{t}(Y)\right], \label{Ko}
\end{equation}
exists for any pair of generators $X,Y\in\frak{sp}(N,\mathbb{R})$
and defines a nilpotent Lie algebra. Moreover, it follows that the
(non-symmetrized) quadratic non-degenerate quadratic Casimir
operator \be
C=2\left(x_{i,-j}x_{-i,j}-x_{i,j}x_{j,i}\right)+x_{i,-i}x_{-i,i}-x_{i,i}^{2}
\ee is preserved by the contraction. In contrast to the previous
example, the contraction has no fixed nilpotency index, and
therefore must not satisfy the sufficient condition (\ref{MA}).

We remark that the contraction method has previously been used in
\cite{Ol} to obtain limits of WZW-models, but without imposing
preservation of invariants by contraction.

\section{Deformations of quasi-classical algebras}\label{sec:three}

Although the obtainment of quasi-classical Lie algebras using
contractions is quite natural, we can also consider a kind of
inverse procedure, namely construct quasi-classical algebras by
deformation of a given one. Like in the preceding case, the Ansatz
is to impose that a non-degenerate quadratic Casimir operator
remains invariant by the deformation. This imposition will lead to
certain tensor equations that the cocycle generating the
deformation must satisfy.

Recall that a cocycle $\varphi\in H^{2}(L,L)$ is a bilinear
skew-symmetric form that satisfies the constraint \cite{Ri}: \bea
d\varphi(X_{i},X_{j},X_{k}):=\left[X_{i},\varphi(X_{j},X_{k})\right]+\left[X_{k},\varphi(X_{i},X_{j})\right]+
\left[X_{j},\varphi(X_{k},X_{i})\right]+\nonumber \\
+\varphi(X_{i},\left[X_{j},X_{k}\right])
+\varphi(X_{k},\left[X_{i},X_{j}\right])+\varphi(X_{j},\left[X_{k},X_{i}\right])=0.
\label{K1} \eea

for all generators $X_{i},X_{j},X_{k}$ of $L$. We further say that
$\varphi$ is integrable if it additionally satisfies the condition
\bea
 \varphi\left(\varphi(X_{i},X_{j}),X_{k}\right)+
\varphi\left(\varphi(X_{j},X_{k}),X_{i}\right)+\varphi\left(\varphi(X_{k},X_{i}),X_{j}\right)=0
\label{K3}.
\eea

Under these conditions, it is straightforward to verify that the
linear deformation $L+\varphi$ is a Lie algebra \cite{Ri} with the
deformed bracket
 \be
\left[X,Y\right]_{\varphi}=\left[X,Y\right]+\varphi(X,Y).
\ee

Supposed that $L$ is a quasi-classical Lie algebra  with
(unsymmetrized) quadratic Casimir operator
$C=g^{\alpha\beta}x_{\alpha}x_{\beta}$, we want to determine
conditions on the integrable cocycle in order to impose that
$L+\varphi$ is also quasi-classical, and has the same quadratic
invariant $C$. To this extent, we realize the deformation
$L+\varphi$ by differential operators, and obtain the system of
PDEs \be
\widehat{X}_{i}=C_{ij}^{k}x_{k}\frac{\partial C}{\partial x_{j}}-\alpha_{ij}^{k}x_{k}%
\frac{\partial C}{\partial x_{j}}=0. \label{L1} \ee

Here
\be
\varphi\left(  X_{i},X_{j}\right)  =\alpha_{ij}^{k}X_{k}%
\ee

is the expression of the cocycle $\varphi$ over the given basis.
Inserting the operator $C$ into the previous system (\ref{L1}), we
obtain \be
\widehat{X}_{i}(C)=C_{ij}^{k}x_{k}g^{\alpha\beta}\frac{\partial\left(
x_{\alpha}x_{\beta }\right)  }{\partial
x_{j}}-\alpha_{ij}^{k}x_{k}g^{\alpha\beta}\frac {\partial\left(
x_{\alpha}x_{\beta}\right)  }{\partial x_{j}},\label{S1} \ee

Since $C$ is an invariant of the undeformed Lie algebra $L$, the
first term reduce to zero, i.e.,
\[
C_{ij}^{k}x_{k}g^{\alpha\beta}\frac{\partial\left(
x_{\alpha}x_{\beta }\right)  }{\partial x_{j}}=0,
\]
and equation (\ref{S1}) reduces to \be \widehat{X}_{i}(C)=-\alpha
_{ij}^{k}x_{k}g^{\alpha\beta}\frac{\partial\left(
x_{\alpha}x_{\beta}\right) }{\partial x_{j}}.\label{S2} \ee

If $C$ is a common invariant of $L$ and $L+\varphi$, then equation
(\ref{S2}) must vanish. Taking into account that for any $1\leq
j\leq N$ the derivatives are given by
\be \frac{\partial}{\partial
x_{j}}\left( g^{kl}x_{k}x_{l}\right)  =\sum_{l\neq
j}g^{lj}x_{l}+2g^{jj}x_{j}, \ee

inserting it into equation (\ref{S2}) and reordering the terms, we
obtain that for any fixed $1\leq i\leq N$ the following system of
equations must be satisfied:

\bea
 \sum_{j=1}^{N}\alpha_{ij}^{i}g^{ij}   =0,\label{T1} \\
 2\alpha_{ij}^{j}g^{jj}+\sum_{k=1}^{N}\alpha_{ik}^{j}g^{ik}+\sum_{k\neq
j}\alpha_{ik}^{i}g^{jk}   =0,\\
2\alpha_{ij}^{j}g^{jj}+\sum_{k\neq j}\alpha_{ik}^{j}g^{kj}   =0,\\
\alpha_{ij}^{j}g^{jk}+\alpha_{ik}^{k}g^{jk}+2\alpha_{ij}^{k}g^{jj}%
+2\alpha_{ik}^{j}g^{kk}+\sum_{l\neq i,j,k}\left(  \alpha_{il}^{j}g^{kl}%
+\alpha_{il}^{k}g^{kl}\right)     =0.\label{T2}
\eea

System (\ref{T1})--(\ref{T2}) provides a necessary and sufficient
condition for a linear deformation $L+\varphi$ (with respect to an
integrable cocycle $\varphi$) to be quasi-classical and preserve
the non-degenerate quadratic Casimir operator of $L$.

As example, consider the indecomposable six dimensional nilpotent
Lie algebra $\frak{n}$ given by the brackets
\[
\left[X_{4},X_{5}\right]=2X_{2},\;
\left[X_{4},X_{6}\right]=-2X_{3},\;
\left[X_{5},X_{6}\right]=-X_{1}.
\]
This algebra is trivially seen to be quasi-classical with
non-degenerate quadratic Casimir operator
$C=x_{1}x_{4}+2x_{2}x_{6}+x_{3}x_{5}$.\footnote{Here the
symetrized form is simply
$C_{symm}=X_{1}X_{4}+2X_{2}X_{6}+X_{3}X_{5}$.} The coefficients of
the associated form are
\[
g^{14}=g^{41}=\frac{1}{2},\; g^{26}=g^{62}=g^{35}=g^{53}=1.
\]
It can be shown that $\dim H^{2}(\frak{n},\frak{n})=30$. Consider
now the nontrivial cocycle given by \bea
\varphi(X_{1},X_{2})=2X_{2},\; \varphi(X_{1},X_{3})=-2X_{3},\;
\varphi(X_{1},X_{5})=2X_{5},\;\varphi(X_{1},X_{6})=-2X_{6},\;\varphi(X_{2},X_{3})=X_{1},\nonumber\\
\varphi(X_{2},X_{4})=-2X_{5},\;\varphi(X_{2},X_{6})=X_{4},\;
\varphi(X_{3},X_{4})=2X_{6},\;\varphi(X_{3},X_{5})=-X_{4}. \eea

It can be verified that $\varphi$ satisfies equation (\ref{K3}),
and is therefore integrable. Let $\frak{g}=\frak{n}+\varphi$ be
the corresponding linear deformation. With some computation it can
be shown that $\varphi$ satisfies the system
(\ref{T1})-(\ref{T2}), which implies that the deformation
$\frak{g}$ is quasi-classical and has $C$ as invariant. Actually,
$\frak{g}$ is isomorphic to the semisimple algebra
$\frak{so}(2,2)$, and considering the maps
\[
f_{t}(X_{i})=\frac{1}{t}X_{i},\; (i=1,2,3),\quad
f_{t}(X_{i})=\frac{1}{\sqrt{t}}X_{i},\; (i=4,5,6)
\]
in $\frak{g}=\frak{n}+\varphi$, the corresponding contraction
recovers $\frak{n}$ and preserves the invariant. Although in
general a deformation is not associated to a contraction
\cite{C63}, this example illustrates a general result, the proof
of which follows by direct computation:

Let $L\rightsquigarrow L^{\prime}$ be a non-trivial contraction of
a quasi-classical Lie algebra $L$ that preserves a non-degenerate
Casimir operator $C$. Then there exists an integrable cocycle
$\varphi\in H^{2}(L^{\prime},L^{\prime})$ such that $\varphi$
satisfies (\ref{T1})-(\ref{T2}) and the linear deformation
$L^{\prime}+\varphi$ preserves the Casimir operator $C$.

We finally remark that deforming quasi-classical Lie algebras by
means of integrable cocycles satisfying conditions
(\ref{T1})-(\ref{T2}) has the advantage of preserving the
signature of the metric. For applications to the WZW model, this
means that the signature of the space-time is preserved, and
therefore, both the deformed and non-deformed models can be
compared since they are described by the same metric. Although a
difficult task in general, it would be an interesting problem to
characterize those Lie algebras which actually admit cocycles of
this type. Work in this direction is in progress.


\section*{Acknowledgements}
\noindent

The author was partially supported by the research project
MTM2006-09152 of the Ministerio de Educaci\'on y Ciencia.


\end{document}